\documentclass[12pt,a4paper]{article}
\usepackage{graphicx}
\usepackage{amsmath}

\begin{document}

\title{Exact vortex solution of the Jacobs-Rebbi equation for ideal fluids}

\author{Florin Spineanu and Madalina Vlad \\
\textit{National Institute for Fusion Science} \\
\textit{322-6 Oroshi-cho, Toki-shi, Gifu-ken, Japan} \\
and \\
\textit{National Institute of Laser, Plasma and Radiation Physics} \\
\textit{P.O.Box MG-36, Magurele, Bucharest, Romania} \\
E-mails: \textit{spineanu@ifin.nipne.ro}, \textit{madi@ifin.nipne.ro}}
\maketitle

\begin{abstract}
The Jacobs-Rebbi equation arises in many contexts where vortical motion in two-dimensional ideal media is investigated. Alternatively, it can be derived in the Abelian Higgs field theory. 
It is considered non-integrable and numerical solutions have been found, consisting of localised, robust
vortices. We show in this work that the equation is integrable and provide the Lax pair. The exact solution is obtained in terms of Riemann \emph{theta} functions.
\end{abstract}

\section{Introduction}

Studying the interaction energy of vortices of the Ginsburg-Landau model of
superconductivity, Jacobs and Rebbi \cite{JacobsRebbi} have derived a nonlinear equation from
which the scalar function (order parameter) can be calculated. The
derivation is based on the similarity of this model with the Abelian Higgs
theory \cite{JackiwPi}, \cite{JacPi90}, \cite{Nardelli}. 
In a special case, corresponding to a particular choice of
parameters, the nonlinear equation takes a simple form, a nonlinear elliptic
differential equation in two spatial dimensions. The same problem is treated
by Dunne \cite{Dunne} in a general context of gauge theories with Maxwell and/or
Chern-Simons terms in the Lagrangean density. The same Abelian Higgs field
theory is discussed and the special case mentioned above is identified as
the minimum of the action functional (saturation of the Bogomolnyi
inequality), realized by fields obeying a simpler set of equations. This
special case is called self-duality, with reference to the equality of the differential two-form
(the curvature of the fibre bundle) with its Hodge dual in the geometrical 
setting of the field-theoretical content.

In fluids and plasmas coherent motion and in particular vortices are
ubiquituous \cite{HH}. They can appear even in turbulent states. In two-dimensions,
apart from the dynamical equations derived from the conservation laws an
alternative model has been proposed, consisting of the motion of discrete,
point-like vortices in plane interacting via a potential \cite{KraichnanMontgomery}. It has been shown
that this model can be mapped onto a field-theoretical model whose structure
is very similar with the non-Abelian gauge-Higgs field theory \cite{FlorinMadi}. The self-dual
state of this field at stationarity is precisely the asymptotic state of the
ideal fluid, which effectively provides an analytic derivation of the \emph{%
sinh}-Poisson equation describing the fluid streamfunction. For the
stationary states attained at very large time by the ideal ion instability
in plasma (described by Hasegawa-Mima), the model of discrete vortices
interacting in plane introduces a short-range potential (then a massive
photon of the gauge field in the field theoretical model).

There are two differences between the field theoretical models developed
starting from the plasma problems and the model from which Jacobs-Rebbi
equation is derived. First, the absolute value of the scalar Higgs field is
constant at large distances (on a circle of very large radius); this means
that the vorticity should be constant at large distance, in the plasma case.
Second, the model must be Abelian, which is less than we would need for
treating the case of the Euler fluid (the \emph{sinh}-Poisson equation).
However, there are physical situations where the boundary conditions for the
vorticity are compatible with the formulation of the Jacobs-Rebbi model.
And, it is known that Abelian models can provide description of plasma
problems (guiding centre particles, for example) leading to the Liouville
equation. The effect of these differences still needs to be investigated,
but in any case, the exact determination of the solution can only be a
useful instrument.

It is usual to consider that the Jacobs-Rebbi equation cannot be solved
analytically and in consequence numerical solutions have been provided.

We show in this paper that the Jacobs-Rebbi equation is exactly integrable.
We consider the integrability on periodic domains and provide the Lax pair.
We follow the standard algebraic-geometric method of integration and
generate explicit solutions in terms of Riemann \emph{theta} functions.

\section{Derivation of the Jacobs-Rebbi equation}

\subsection{Derivation in the context of Ginsburg-Landau theory}

The Ginsburg-Landau theory is the framework in which the Jacobs-Rebbi equation has been derived,
since the original aim was the investigation of the energy of interaction
between two vortices in superconducting Helium. The interest for vortical
structures of Ginsburg-Landau comes also from the observation that the
theory of a gauge field coupled to a scalar field (Abelian Higgs field) can
in some cases exhibit also coherent vortical structures. We include in this Section the
derivation according by Jacobs and Rebbi \cite{JacobsRebbi}. In fluid physics
other approaches can be developed to arrive at similar forms of the equation.

The free energy of the Ginsburg-Landau theory and the potential energy of
the Abelian-Higgs theory is 
\begin{eqnarray*}
E &=&\int d^{3}x\left\{ \frac{1}{4}F_{ij}F^{ij}\right. \\
&&\left. +\frac{1}{2}\left| \left( \partial _{i}-ieA_{i}\right) \phi \right|
^{2}+c_{4}\left( \left| \phi \right| ^{2}-c_{0}^{2}\right) ^{2}\right\}
\end{eqnarray*}
where $\phi $ is a complex scalar field, $A_{i}$ is the Abelian gauge
potential and 
\begin{equation*}
F_{ij}=\partial _{i}A_{j}-\partial _{j}A_{i}
\end{equation*}
The minimum of the energy is attained for 
\begin{equation*}
\left| \phi \right| =c_{0}\neq 0
\end{equation*}

The variables are rescaled as 
\begin{equation*}
x_{i}=\frac{1}{c_{0}e}\widetilde{x}_{i}
\end{equation*}
\begin{equation*}
A_{i}=c_{0}\widetilde{A}_{i}
\end{equation*}
\begin{equation*}
\phi =c_{0}\widetilde{\phi }
\end{equation*}
and the energy becoms 
\begin{eqnarray*}
E &=&\frac{c_{0}}{e}\int d^{3}\widetilde{x}\left\{ \frac{1}{4}\widetilde{F}%
_{ij}\widetilde{F}^{ij}\right. \\
&&\left. +\frac{1}{2}\left| \left( \widetilde{\partial }_{i}-ie\widetilde{A}%
_{i}\right) \widetilde{\phi }\right| ^{2}+\frac{\lambda ^{2}}{8}\left(
\left| \widetilde{\phi }\right| ^{2}-1\right) ^{2}\right\}
\end{eqnarray*}
where 
\begin{equation*}
\lambda ^{2}=\frac{8c_{4}}{e^{2}}
\end{equation*}

The model is restricted to the case where all field functions does not
depend on the third coordinate and 
\begin{equation*}
A_{3}\equiv 0
\end{equation*}
In plane the coordinates are expressed by complex variables 
\begin{eqnarray*}
z &=&\widetilde{x}_{1}+i\widetilde{x}_{2} \\
\overline{z} &=&\widetilde{x}_{1}-i\widetilde{x}_{2}
\end{eqnarray*}
and the differential operators can be defined 
\begin{eqnarray*}
\partial &\equiv &\frac{1}{2}\left( \frac{\partial }{\partial \widetilde{x}%
_{1}}-i\frac{\partial }{\partial \widetilde{x}_{2}}\right) \\
\overline{\partial } &\equiv &\frac{1}{2}\left( \frac{\partial }{\partial 
\widetilde{x}_{1}}+i\frac{\partial }{\partial \widetilde{x}_{2}}\right)
\end{eqnarray*}
Analoguous combinations are used for the two remaining potential components 
\begin{eqnarray*}
A &=&\frac{1}{2}\left( \widetilde{A}_{1}-i\widetilde{A}_{2}\right) \\
\overline{A} &=&\frac{1}{2}\left( \widetilde{A}_{1}+i\widetilde{A}_{2}\right)
\end{eqnarray*}

The energy per unit length along the coordinate axis $x_{3}$ is 
\begin{equation*}
E=\frac{c_{0}\pi }{e}\mathcal{E}
\end{equation*}
\begin{eqnarray*}
\mathcal{E} &=&\frac{1}{2\pi }\int dzd\overline{z}\left\{ 2\left| \overline{%
\partial }A-\partial \overline{A}\right| \right. \\
&&\left. +\left| \left( \partial -iA\right) \widetilde{\phi }\right|
^{2}+\left| \left( \overline{\partial }-i\overline{A}\right) \widetilde{\phi 
}\right| ^{2}+\frac{\lambda ^{2}}{8}\left( \left| \widetilde{\phi }\right|
^{2}-1\right) ^{2}\right\}
\end{eqnarray*}

In the following the \emph{tilda} will be omitted.

The equations of motion that are derived from the Lagrangean 
\begin{equation*}
\left( \partial -iA\right) \left( \overline{\partial }-i\overline{A}\right)
\phi +\left( \overline{\partial }-i\overline{A}\right) \left( \partial
-iA\right) \phi -\frac{\lambda ^{2}}{4}\phi \left( \left| \phi \right|
^{2}-1\right) =0
\end{equation*}
\begin{eqnarray*}
&&4\partial \overline{\partial }A-4\partial ^{2}A \\
&&-i\overline{\phi }\partial \phi +i\phi \partial \overline{\phi } \\
&&-2A\phi \overline{\phi } \\
&=&0
\end{eqnarray*}

For a general value of $\lambda $ one can replace particular forms for the
two functions $\phi $ and $A$ and obtain differential equations.

For the value 
\begin{equation*}
\lambda =1
\end{equation*}
the situation is different since one can obtain a lower bound for the energy.

The following integration by parts is done 
\begin{eqnarray*}
&&\int dzd\overline{z}\left[ \left( \partial -iA\right) \phi \left( 
\overline{\partial }+i\overline{A}\right) \overline{\phi }\right]  \\
&=&\int dzd\overline{z}\left[ \left( \overline{\partial }-i\overline{A}%
\right) \overline{\phi }\left( \partial +iA\right) \phi \right.  \\
&&\hspace*{1cm}\left. -i\left( \overline{\partial }A-\partial \overline{A}%
\right) \phi \overline{\phi }\right] 
\end{eqnarray*}
Then one obtains the new expression for the energy 
\begin{eqnarray*}
\mathcal{E} &=&\frac{1}{\pi }\int dzd\overline{z}\left\{ \left| \left( 
\overline{\partial }-i\overline{A}\right) \phi \right| ^{2}+\left[ -i\left( 
\overline{\partial }A-\partial \overline{A}\right) -\frac{1}{4}\left( \left|
\phi \right| ^{2}-1\right) \right] ^{2}\right\}  \\
&&-\frac{i}{2\pi }\int dzd\overline{z}\left( \overline{\partial }A-\partial 
\overline{A}\right) 
\end{eqnarray*}
Taking into account the boundary conditions and asking for the absolute
minimum to be attained the terms in the curly braket must be taken zero and
we obtain the equations 
\begin{equation*}
\left( \overline{\partial }-i\overline{A}\right) \phi =0
\end{equation*}
\begin{equation*}
\overline{\partial }A-\partial \overline{A}+\frac{i}{4}\left( \left| \phi
\right| ^{2}-1\right) =0
\end{equation*}
This equations can be further transformed 
\begin{eqnarray*}
A &=&i\partial \psi  \\
\overline{A} &=&-i\overline{\partial }\psi 
\end{eqnarray*}
then the first equation reduces to 
\begin{equation*}
\left( \overline{\partial }-\overline{\partial }\psi \right) \phi =\exp
\left( \psi \right) \overline{\partial }\left[ \exp \left( -\psi \right)
\phi \right] =0
\end{equation*}
which means that we can introduce an analytic function 
\begin{equation*}
f=\exp \left( -\psi \right) \phi 
\end{equation*}
Inserting 
\begin{equation*}
\phi \left( z,\overline{z}\right) =\exp \left[ -\psi \left( z,\overline{z}%
\right) \right] f\left( z\right) 
\end{equation*}
in the second equation we obtain 
\begin{equation*}
\partial \overline{\partial }\psi =\frac{1}{8}\left[ \exp \left( 2\psi
\right) f\overline{f}-1\right] 
\end{equation*}
By a new substitution 
\begin{equation*}
\psi =\chi -\frac{1}{2}\ln \left( f\overline{f}\right) 
\end{equation*}
the equation becomes 
\begin{equation*}
\partial \overline{\partial }\chi =\frac{1}{8}\left[ \exp \left( 2\chi
\right) -1\right] 
\end{equation*}
with the condition the $\chi $ goes to zero at infinity.

\section{General procedure for obtaining solutions of the Jacobs-Rebbi
equation}

The procedure is similar to those developed for the \emph{sine}-Gordon equation \cite{ForestMcLaughlin}
and for sinh-Poisson equation \cite{TingChenLee}.
Just as in the general case of nonlinear differential equations which are
exctly integrable by the algebraic-geometric procedure, we start from a
configuration which is specified initially. By contrast with other equations
(for example KdV, etc) where the variables are space and time and the
unknown function is given at $t=0$, here the coordinates are both spatial.
Then the conditions to be specified are boundary conditions, for example
taken on the lines $x=0$ and $y=0$.

Consider that a certain flow configuration is specified (for example from
experimental measurements) and the boundary conditions are specified in the
form of two functions 
\begin{equation*}
u_{0x}\left( y\right) \;\text{and}\;u_{0y}\left( x\right)
\end{equation*}
for $x\in \left[ 0,L\right] $ and $y\in \left[ 0,L\right] $. We assume that $%
L$ is much smaller than the radius of the circle on which the asymptotic
value of the vorticity is given. The ``initial'' values of the unknown
function is introduced in the Lax operator eigenvalue problem. Solving this
problem we identify a set of eigenvalues (the Lax operator spectrum, see \cite{Ablowitz}) and the
corresponding eigenfunctions with periodicity properties (Bloch functions).
It is a general situation that in the spectrum there is a subset of
eigenvalues for which the two eigenfunctions are identical. These
eigenvalues are called \emph{non-degenerate} and the subset is called \emph{%
main spectrum}.

Using the main spectrum one can construct the hyperelliptic Riemann surface
associated with the Wronskian of the eigenfunctions. For a two by two Lax
operator, (i.e. hyperelliptic Riemann surface) the monodromy problem is
simple.

One has to define on this surface the dual homological sets: cycles and
differential one-forms. Then the period matrices can be calculated.

Using the inverse of the $A$-period matrix one can generate the variables
(the phases) appearing in the arguments of the Riemann \emph{theta} function.

Finally, one can calculate the solution at any point $\left( x,y\right) $
and can represent it graphically on a space domain.

\subsection{The spectral problem for the Jacobi-Rebbi equation on periodic
domain}

From a detailed consideration of Lax pairs found by Forest and McLaughlin
for the \emph{sine}-Gordon equation \cite{ForestMcLaughlin}, we obtain the following Lax equations.
The first is 
\begin{equation*}
\left( 
\begin{array}{cc}
-\frac{\lambda ^{2}}{16\sqrt{p}}-\sqrt{p} & -i\frac{\partial }{\partial x}-%
\frac{1}{4}\left( \frac{\partial u}{\partial y}+i\frac{\partial u}{\partial x%
}\right)  \\ 
i\frac{\partial }{\partial x}-\frac{1}{4}\left( \frac{\partial u}{\partial y}%
+i\frac{\partial u}{\partial x}\right)  & -\frac{\lambda ^{2}}{16\sqrt{p}}%
\exp \left( u\right) -\sqrt{p}
\end{array}
\right) \left( 
\begin{array}{c}
\psi _{1} \\ 
\psi _{2}
\end{array}
\right) =0
\end{equation*}
This set of equations is considered on a periodic domain along the $x$ axis.
It is of second differential order and has two independent solutions
periodic on $x$, which we note $\phi _{+}$ and $\phi _{-}$. We chose them to
correspond to the following initial conditions at $x=x_{0}$%
\begin{eqnarray*}
\phi _{+}\left( x_{0},x_{0},p\right)  &\equiv &\left( 
\begin{array}{c}
1 \\ 
0
\end{array}
\right)  \\
\phi _{-}\left( x_{0},x_{0},p\right)  &\equiv &\left( 
\begin{array}{c}
0 \\ 
1
\end{array}
\right) 
\end{eqnarray*}
Any other solution of the system, corresponding to the following condition
taken at $x=x_{0}$ 
\begin{equation*}
\phi \left( x=x_{0},p\right) =\left( 
\begin{array}{c}
P \\ 
Q
\end{array}
\right) 
\end{equation*}
is a linear combination of these two basis functions 
\begin{equation*}
\phi \left( x,x_{0},p\right) \equiv \left( 
\begin{array}{c}
\phi _{1}\left( x,x_{0},p\right)  \\ 
\phi _{2}\left( x,x_{0},p\right) 
\end{array}
\right) =P\phi _{+}\left( x,x_{0},p\right) +Q\phi _{-}\left(
x,x_{0},p\right) 
\end{equation*}

We consider the second set of equations 
\begin{equation*}
\left( 
\begin{array}{cc}
\frac{\lambda ^{2}}{16\sqrt{p}}-\sqrt{p} & -\frac{\partial }{\partial y}-%
\frac{1}{4}\left( \frac{\partial u}{\partial y}+i\frac{\partial u}{\partial x%
}\right) \\ 
\frac{\partial }{\partial y}-\frac{1}{4}\left( \frac{\partial u}{\partial y}%
+i\frac{\partial u}{\partial x}\right) & \frac{\lambda ^{2}}{16\sqrt{p}}\exp
\left( u\right) -\sqrt{p}
\end{array}
\right) \left( 
\begin{array}{c}
\psi _{1} \\ 
\psi _{2}
\end{array}
\right) =0
\end{equation*}
This is a system of two differential equations on the periodic domain along
the $y$ axis, having two independent solutions. These two functions must
actually be identical to the previously defined functions, $\phi _{+}$ and $%
\phi _{-}$ since finally we can only accept solutions of both sets of
equations, on $x$ and on $y$. On the $y$ direction we take the initial
conditions 
\begin{eqnarray*}
\phi _{+}\left( y_{0},y_{0},p\right) &=&\left( 
\begin{array}{c}
1 \\ 
0
\end{array}
\right) \\
\phi _{-}\left( y_{0},y_{0},p\right) &=&\left( 
\begin{array}{c}
0 \\ 
1
\end{array}
\right)
\end{eqnarray*}
Any other solution of the system, corresponding to the following condition
taken at $y=y_{0}$ 
\begin{equation*}
\psi \left( y=y_{0},p\right) =\left( 
\begin{array}{c}
P^{\prime } \\ 
Q^{\prime }
\end{array}
\right)
\end{equation*}
is a linear combination of these two basis functions 
\begin{equation*}
\psi \left( y,y_{0},p\right) \equiv \left( 
\begin{array}{c}
\psi _{1}\left( y,y_{0},p\right) \\ 
\psi _{2}\left( y,y_{0},p\right)
\end{array}
\right) =P^{\prime }\phi _{+}\left( y,y_{0},p\right) +Q^{\prime }\phi
_{-}\left( y,y_{0},p\right)
\end{equation*}
The fact that we write only the $x$ or the $y$ notation is only derived from
the context of the first or second system. Actually the pair of functions $%
\left( \phi ,\psi \right) $ depend on $\left( x,y\right) $ on a
two-dimensional periodic domain and are independent solutions of the two
systems of equations. We note 
\begin{equation*}
w=i\left( \frac{\partial u}{\partial y}+i\frac{\partial u}{\partial x}\right)
\end{equation*}
\begin{equation*}
\left( 
\begin{array}{cc}
-\frac{\lambda ^{2}}{16\sqrt{p}}-\sqrt{p} & -i\frac{\partial }{\partial x}+%
\frac{i}{4}w \\ 
i\frac{\partial }{\partial x}+\frac{i}{4}w & -\frac{\lambda ^{2}}{16\sqrt{p}}%
\exp \left( u\right) -\sqrt{p}
\end{array}
\right) \left( 
\begin{array}{c}
\phi _{1} \\ 
\phi _{2}
\end{array}
\right) =0
\end{equation*}
\begin{equation*}
\left( 
\begin{array}{cc}
\frac{\lambda ^{2}}{16\sqrt{p}}-\sqrt{p} & -\frac{\partial }{\partial y}+%
\frac{i}{4}w \\ 
\frac{\partial }{\partial y}+\frac{i}{4}w & \frac{\lambda ^{2}}{16\sqrt{p}}%
\exp \left( u\right) -\sqrt{p}
\end{array}
\right) \left( 
\begin{array}{c}
\psi _{1} \\ 
\psi _{2}
\end{array}
\right) =0
\end{equation*}
or 
\begin{eqnarray*}
\left( -\frac{\lambda ^{2}}{16\sqrt{p}}-\sqrt{p}\right) \phi _{1}-i\frac{%
\partial \phi _{2}}{\partial x}+\frac{i}{4}w\phi _{2} &=&0 \\
i\frac{\partial \phi _{1}}{\partial x}+\frac{i}{4}w\phi _{1}+\left[ -\frac{%
\lambda ^{2}}{16\sqrt{p}}\exp \left( u\right) -\sqrt{p}\right] \phi _{2} &=&0
\end{eqnarray*}
\begin{eqnarray*}
\left( \frac{\lambda ^{2}}{16\sqrt{p}}-\sqrt{p}\right) \phi _{1}-\frac{%
\partial \phi _{2}}{\partial y}+\frac{i}{4}w\phi _{2} &=&0 \\
\frac{\partial \phi _{1}}{\partial y}+\frac{i}{4}w\phi _{1}+\left[ \frac{%
\lambda ^{2}}{16\sqrt{p}}\exp \left( u\right) -\sqrt{p}\right] \phi _{2} &=&0
\end{eqnarray*}
\begin{eqnarray*}
\left( -\frac{\lambda ^{2}}{16\sqrt{p}}-\sqrt{p}\right) \psi _{1}-i\frac{%
\partial \psi _{2}}{\partial x}+\frac{i}{4}w\psi _{2} &=&0 \\
i\frac{\partial \psi _{1}}{\partial x}+\frac{i}{4}w\psi _{1}+\left[ -\frac{%
\lambda ^{2}}{16\sqrt{p}}\exp \left( u\right) -\sqrt{p}\right] \psi _{2} &=&0
\end{eqnarray*}
\begin{eqnarray*}
\left( \frac{\lambda ^{2}}{16\sqrt{p}}-\sqrt{p}\right) \psi _{1}-\frac{%
\partial \psi _{2}}{\partial y}+\frac{i}{4}w\psi _{2} &=&0 \\
\frac{\partial \psi _{1}}{\partial y}+\frac{i}{4}w\psi _{1}+\left[ \frac{%
\lambda ^{2}}{16\sqrt{p}}\exp \left( u\right) -\sqrt{p}\right] \psi _{2} &=&0
\end{eqnarray*}

According to standard procedures we define squared eigenfunctions, as
combinations of the components of the two independent solutions $\phi $ and $%
\psi $. 
\begin{eqnarray}
f &=&-\frac{i}{2}\left( \psi _{1}\phi _{2}+\phi _{1}\psi _{2}\right)
\label{fgh} \\
g &=&\psi _{1}\phi _{1}  \notag \\
h &=&-\psi _{2}\phi _{2}  \notag
\end{eqnarray}
and calculate the derivatives at $x$ and at $y$. 
\begin{eqnarray}
\frac{\partial f}{\partial x} &=&\frac{1}{2}\left( \sqrt{p}+\frac{\lambda
^{2}}{16\sqrt{p}}\right) g+\frac{1}{2}\left[ \sqrt{p}+\frac{\lambda ^{2}}{16%
\sqrt{p}}\exp \left( u\right) \right] h  \label{fghx} \\
\frac{\partial g}{\partial x} &=&2\left[ \frac{\lambda ^{2}}{16\sqrt{p}}\exp
\left( u\right) +\sqrt{p}\right] f-\frac{w}{2}g  \notag \\
\frac{\partial h}{\partial x} &=&2\left( \frac{\lambda ^{2}}{16\sqrt{p}}+%
\sqrt{p}\right) f+\frac{w}{2}h  \notag
\end{eqnarray}
and 
\begin{eqnarray}
\frac{\partial f}{\partial y} &=&i\left[ \sqrt{p}-\frac{\lambda ^{2}}{16%
\sqrt{p}}\exp \left( u\right) \right] h+i\left( \sqrt{p}-\frac{\lambda ^{2}}{%
16\sqrt{p}}\right) g  \label{fghy} \\
\frac{\partial g}{\partial y} &=&-\frac{iw}{2}g+2i\left[ \frac{\lambda ^{2}}{%
16\sqrt{p}}\exp \left( u\right) -\sqrt{p}\right] f  \notag \\
\frac{\partial h}{\partial y} &=&\frac{iw}{2}h+2i\left( \frac{\lambda ^{2}}{%
16\sqrt{p}}-\sqrt{p}\right) f  \notag
\end{eqnarray}
Using the squared eigenfunction it is possible to construct the constant of
motion 
\begin{equation}
C=f^{2}-gh  \label{c}
\end{equation}
with the properties 
\begin{eqnarray*}
\frac{\partial C}{\partial x} &=&0 \\
\frac{\partial C}{\partial y} &=&0
\end{eqnarray*}
Then $C$ depends only on the eigenvalue $p$%
\begin{equation*}
C\equiv C\left( p\right)
\end{equation*}
It can be shown that the Wronskian of two solutions of the systems of
equations 
\begin{eqnarray*}
W &=&\det \left( 
\begin{array}{cc}
\phi _{1} & \psi _{1} \\ 
\phi _{2} & \psi _{2}
\end{array}
\right) \\
&=&\phi _{1}\psi _{2}-\phi _{2}\psi _{1}
\end{eqnarray*}
can be expressed in terms of this constant of motion by the relation 
\begin{equation*}
C=-\frac{W^{2}}{4}
\end{equation*}

The fact that we have a formal expression for the Wronskian in terms of a
function of only the eigenvalue $p$, allows us to discuss the problem of the
existence of two independent solutions to the systems of equations. There
will be independent solutions everywhere on the complex $p$ plane except at
the points where the Wronskian vanishes. The set of points on the complex $p$
plane where the Wronskian vanishes (and there is only one solution) is
called \emph{main spectrum} of the scattering problem.

We will write the squared Wronskian (\emph{i.e.} $C$) as a polynomial of the
variable $p$ thus formally introducing the points of the main spectrum, $%
p_{i},i=1,2N$. 
\begin{equation}
-\frac{1}{4}W^{2}=\prod_{i=1}^{2N}\left( p-p_{i}\right)  \label{Wpoly}
\end{equation}

Since there is a relation between the squared eigenfunctions and the
Wronskian, we will introduce analoguous expressions as polynomial in the
variable $p$ 
\begin{eqnarray}
f &=&\frac{1}{\sqrt{p}}\sum_{k=1}^{N}f_{k}p^{k}  \label{fghpoly} \\
g &=&\sum_{k=0}^{N}g_{k}p^{k}  \notag \\
h &=&\sum_{k=0}^{N}h_{k}p^{k}  \notag
\end{eqnarray}
where the coefficients are functions of $\left( x,y\right) $.

We dispose of differential equations relating these functions, Eqs.(\ref
{fghx}) and (\ref{fghy}), and we will insert the polynomial expansion and
find relations between the coefficients. 
\begin{eqnarray*}
\frac{1}{\sqrt{p}}\sum_{k=1}^{N}\frac{\partial f_{k}}{\partial x}p^{k} &=&%
\frac{1}{2}\sqrt{p}\sum_{k=0}^{N}\left( g_{k}+h_{k}\right) p^{k} \\
&&+\frac{1}{2}\frac{\lambda ^{2}}{16\sqrt{p}}\sum_{k=0}^{N}\left[ g_{k}+\exp
\left( u\right) h_{k}\right] p^{k}
\end{eqnarray*}
It results 
\begin{equation}
0=\frac{1}{2}\frac{\lambda ^{2}}{16}\left[ g_{0}+h_{0}\exp \left( u\right) %
\right]  \label{eq1}
\end{equation}
\begin{equation*}
\frac{\partial f_{1}}{\partial x}=\frac{1}{2}\left( g_{0}+h_{0}\right) +%
\frac{1}{2}\frac{\lambda ^{2}}{16}\left[ g_{1}+h_{1}\exp \left( u\right) %
\right]
\end{equation*}
\begin{equation}
\frac{\partial f_{k}}{\partial x}=\frac{1}{2}\left( g_{k-1}+h_{k-1}\right) +%
\frac{1}{2}\frac{\lambda ^{2}}{16}\left[ g_{k}+h_{k}\exp \left( u\right) %
\right] \;,\;k=2,...,N  \label{eq2}
\end{equation}
\begin{equation}
0=\frac{1}{2}\left( g_{N}+h_{N}\right)  \label{eq3}
\end{equation}

Using now the definition of the Wronskian and the polynomial expressions 
\begin{eqnarray}
C &=&f^{2}-gh  \label{ccapol} \\
&=&\left( \frac{1}{\sqrt{p}}\sum_{k=1}^{N}f_{k}p^{k}\right)
^{2}-\sum_{k=0}^{N}g_{k}p^{k}\sum_{k=0}^{N}h_{k}p^{k}  \notag \\
&=&\prod_{k=1}^{2N}\left( p-p_{k}\right)  \notag
\end{eqnarray}
In order the invariant quantity $C\left( p\right) $ (or the Wronskian) to be
a polynomial in $p$, there should be no source of singularity in Eq.(\ref
{ccapol}) and this means that the function $f\left( p\right) $ must have 
\begin{equation*}
f_{0}\equiv 0
\end{equation*}
as we have already taken in (\ref{fghx}).

The coefficient of the zero-degree of $p$ in $C\left( p\right) $ is 
\begin{equation}
C\left( p=0\right) =-g_{0}h_{0}  \label{g0h0}
\end{equation}

Eq.(\ref{ccapol}) gives 
\begin{equation}
C\left( 0\right) =\left( -1\right) ^{2N}\prod_{k=1}^{2N}p_{k}=-g_{0}h_{0}
\label{czero}
\end{equation}

We now introduce the \emph{zeros} $\gamma _{k}\left( x,y\right) ,k=1,...,N$
of the function $g\left( x,y;p\right) $ (we suppress the arguments $\left(
x,y\right) $) 
\begin{equation}
g\left( p\right) =\prod_{k=1}^{N}\left( p-\gamma _{k}\right)  \label{ggamma}
\end{equation}
from which it results 
\begin{equation}
g\left( 0\right) =\left( -1\right) ^{N}\prod_{k=1}^{N}\gamma _{k}=g_{0}
\label{gzero}
\end{equation}
From the equations (\ref{czero}), (\ref{g0onh0}) and (\ref{gzero}) we obtain 
\begin{equation}
\exp \left( u\right) =-\frac{g_{0}^{2}}{h_{0}g_{0}}=\frac{\left[ \left(
-1\right) ^{N}\prod_{k=1}^{N}\gamma _{k}\right] ^{2}}{\prod_{k=1}^{2N}p_{k}}
\label{expugp}
\end{equation}
or 
\begin{equation}
u=\ln \left[ \frac{\left( \prod_{k=1}^{N}\gamma _{k}\right) ^{2}}{%
\prod_{k=1}^{2N}p_{k}}\right]  \label{expu}
\end{equation}

Eq.(\ref{expu}) shows that if we have the \emph{main spectrum} and if we
could calculate the \emph{zeros of the squared eigenfunction} $g$, we could
find the solution to the nonlinear equation. The ensemble of zeros of the
squared eigenfunction $g$ is called \emph{auxiliary spectrum}.

\subsection{The equations for the auxiliary spectrum}

To find the differential equations obeyed by $\gamma _{k}$ we start from the
Eqs.(\ref{fghx}) for $g$ 
\begin{equation}
\frac{\partial g}{\partial x}=2\left[ \frac{\lambda ^{2}}{16\sqrt{p}}\exp
\left( u\right) +\sqrt{p}\right] f-\frac{w}{2}g  \label{glax}
\end{equation}
and its $y$ version 
\begin{equation}
\frac{\partial g}{\partial y}=-\frac{iw}{2}g+2i\left[ \frac{\lambda ^{2}}{16%
\sqrt{p}}\exp \left( u\right) -\sqrt{p}\right] f  \label{glay}
\end{equation}
and calculate all terms at $p=\gamma _{k}$, a zero of $g$, using Eq.(\ref
{ggamma}) 
\begin{eqnarray}
\left. \frac{\partial g}{\partial x}\right| _{p=\gamma _{k}} &=&-\frac{%
\partial \gamma _{k}}{\partial x}\prod_{\substack{ l=1  \\ l\neq k}}%
^{N}\left( \gamma _{k}-\gamma _{l}\right)  \label{dgx1} \\
&=&2\left[ \frac{\lambda ^{2}}{16\sqrt{\gamma _{k}}}\exp \left( u\right) +%
\sqrt{\gamma _{k}}\right] f\left( \gamma _{k}\right)  \notag
\end{eqnarray}
The value of $f\left( \gamma _{k}\right) $ is determined from the expression
of $C\left( p\right) $ , Eqs.(\ref{c}) and (\ref{ccapol}), after inserting $%
p=\gamma _{k}$%
\begin{equation*}
\prod_{l=1}^{2N}\left( \gamma _{k}-p_{l}\right) =\left[ f\left( \gamma
_{k}\right) \right] ^{2}
\end{equation*}
In Eq.(\ref{dgx1}) we will also replace $\exp \left( u\right) $ from Eq.(\ref
{expugp}). Then 
\begin{equation}
\frac{\partial \gamma _{k}}{\partial x}=-2\left[ \frac{\lambda ^{2}}{16\sqrt{%
\gamma _{k}}}\frac{\left( \prod_{l=1}^{N}\gamma _{l}\right) ^{2}}{%
\prod_{l=1}^{2N}p_{l}}+\sqrt{\gamma _{k}}\right] \frac{\left[
\prod_{l=1}^{2N}\left( \gamma _{k}-p_{l}\right) \right] ^{1/2}}{\prod 
_{\substack{ l=1  \\ l\neq k}}^{N}\left( \gamma _{k}-\gamma _{l}\right) }
\label{dgammax}
\end{equation}
In a similar way we have 
\begin{eqnarray*}
\left. \frac{\partial g}{\partial y}\right| _{p=\gamma _{k}} &=&-\frac{%
\partial \gamma _{k}}{\partial y}\prod_{\substack{ l=1  \\ l\neq k}}%
^{N}\left( \gamma _{k}-\gamma _{l}\right) \\
&=&2i\left[ \frac{\lambda ^{2}}{16\sqrt{\gamma _{k}}}\exp \left( u\right) -%
\sqrt{\gamma _{k}}\right] f\left( \gamma _{k}\right)
\end{eqnarray*}
or 
\begin{equation}
\frac{\partial \gamma _{k}}{\partial y}=-2i\left[ \frac{\lambda ^{2}}{16%
\sqrt{\gamma _{k}}}\frac{\left( \prod_{l=1}^{N}\gamma _{l}\right) ^{2}}{%
\prod_{l=1}^{2N}p_{l}}-\sqrt{\gamma _{k}}\right] \frac{\left[
\prod_{l=1}^{2N}\left( \gamma _{k}-p_{l}\right) \right] ^{1/2}}{\prod 
_{\substack{ l=1  \\ l\neq k}}^{N}\left( \gamma _{k}-\gamma _{l}\right) }
\label{dgammay}
\end{equation}

\subsection{Checking the formulas as solutions}

As Ting, Chen, Lee \cite{TingChenLee} have shown for the case of the \emph{sinh}-Poisson
equation, it may be useful to try to find out if the formulas determined
above, Eqs.(\ref{dgammax}) and (\ref{dgammay}) may already be taken as
solution, for a set of $\gamma _{k}$ which is not yet determined. The
procedure consists of replacing the expression (\ref{expu}) in the initial
equation, perform the derivatives of the functions $\gamma _{k}\left(
x,y\right) $ appearing in this expression and taking into account the
equations of motion, Eqs.(\ref{dgammax}) and (\ref{dgammay}).

The following change of variables makes the calculation easier 
\begin{equation*}
x\rightarrow x^{\prime }=ix+y
\end{equation*}
\begin{equation*}
y\rightarrow y^{\prime }=-ix+y
\end{equation*}
and the initial equation becomes 
\begin{equation*}
4\frac{\partial ^{2}u}{\partial x^{\prime }\partial y^{\prime }}=\frac{%
\lambda ^{2}}{2}\left[ \exp \left( u\right) -1\right]
\end{equation*}

The equations of motions are translated in the new variables 
\begin{eqnarray*}
&&\frac{\partial \gamma _{k}}{\partial x^{\prime }}-\frac{\partial \gamma
_{k}}{\partial y^{\prime }} \\
&=&2i\left[ \frac{\lambda ^{2}}{16\sqrt{\gamma _{k}}}\frac{\left(
\prod_{l=1}^{N}\gamma _{l}\right) ^{2}}{\prod_{l=1}^{2N}p_{l}}+\sqrt{\gamma
_{k}}\right] \frac{\left[ \prod_{l=1}^{2N}\left( \gamma _{k}-p_{l}\right) %
\right] ^{1/2}}{\prod_{\substack{ l=1  \\ l\neq k}}^{N}\left( \gamma
_{k}-\gamma _{l}\right) }
\end{eqnarray*}
\begin{eqnarray*}
&&\frac{\partial \gamma _{k}}{\partial x^{\prime }}+\frac{\partial \gamma
_{k}}{\partial y^{\prime }} \\
&=&-2i\left[ \frac{\lambda ^{2}}{16\sqrt{\gamma _{k}}}\frac{\left(
\prod_{l=1}^{N}\gamma _{l}\right) ^{2}}{\prod_{l=1}^{2N}p_{l}}-\sqrt{\gamma
_{k}}\right] \frac{\left[ \prod_{l=1}^{2N}\left( \gamma _{k}-p_{l}\right) %
\right] ^{1/2}}{\prod_{\substack{ l=1  \\ l\neq k}}^{N}\left( \gamma
_{k}-\gamma _{l}\right) }
\end{eqnarray*}
Adding and substracting these equations we obtain 
\begin{equation}
\frac{\partial \gamma _{k}}{\partial x^{\prime }}=2i\sqrt{\gamma _{k}}\frac{%
\left[ \prod_{l=1}^{2N}\left( \gamma _{k}-p_{l}\right) \right] ^{1/2}}{\prod 
_{\substack{ l=1  \\ l\neq k}}^{N}\left( \gamma _{k}-\gamma _{l}\right) }
\label{dgdxp}
\end{equation}
\begin{equation*}
\frac{\partial \gamma _{k}}{\partial y^{\prime }}=-i\frac{\lambda ^{2}}{8%
\sqrt{\gamma _{k}}}\frac{\left( \prod_{l=1}^{N}\gamma _{l}\right) ^{2}}{%
\prod_{l=1}^{2N}p_{l}}\frac{\left[ \prod_{l=1}^{2N}\left( \gamma
_{k}-p_{l}\right) \right] ^{1/2}}{\prod_{\substack{ l=1  \\ l\neq k}}%
^{N}\left( \gamma _{k}-\gamma _{l}\right) }
\end{equation*}

This last expression can be written, using Eqs.(\ref{expugp}) and (\ref
{dgdxp}) 
\begin{equation}
\frac{\partial \gamma _{k}}{\partial y^{\prime }}=-\frac{\lambda ^{2}}{16}%
\frac{1}{\gamma _{k}}\exp \left( u\right) \frac{\partial \gamma _{k}}{%
\partial x^{\prime }}  \label{dgdyp}
\end{equation}

The conversion formula is 
\begin{equation}
\exp \left( u\right) =\frac{\left( \prod_{l=1}^{N}\gamma _{l}\right) ^{2}}{%
\prod_{l=1}^{2N}p_{l}}  \label{conv}
\end{equation}
and can be used to obtain the derivatives of $u$%
\begin{eqnarray*}
\frac{\partial u}{\partial x^{\prime }} &=&2\exp \left( -u\right) \frac{%
\left( \prod_{l=1}^{N}\gamma _{l}\right) ^{2}}{\prod_{l=1}^{2N}p_{l}}%
\sum_{l=1}^{N}\frac{1}{\gamma _{l}}\frac{\partial \gamma _{l}}{\partial
x^{\prime }} \\
&=&4i\sum_{l=1}^{N}\frac{1}{\sqrt{\gamma _{l}}}\frac{\left[
\prod_{m=1}^{2N}\left( \gamma _{l}-p_{m}\right) \right] ^{1/2}}{\prod 
_{\substack{ m=1  \\ m\neq l}}^{N}\left( \gamma _{l}-\gamma _{m}\right) }
\end{eqnarray*}
\begin{eqnarray*}
\frac{\partial ^{2}u}{\partial y^{\prime }\partial x^{\prime }}
&=&4i\sum_{l=1}^{N}\frac{1}{\sqrt{\gamma _{l}}}\left\{ \frac{1}{\prod 
_{\substack{ m=1  \\ m\neq l}}^{N}\left( \gamma _{l}-\gamma _{m}\right) }%
\right. \\
&&\times \frac{1}{2}\left[ \prod_{m=1}^{2N}\left( \gamma _{l}-p_{m}\right) %
\right] ^{1/2}\sum_{m=1}^{2N}\frac{1}{\left( \gamma _{l}-p_{m}\right) }\frac{%
\partial \gamma _{l}}{\partial y^{\prime }} \\
&&\left. -\frac{\left[ \prod_{m=1}^{2N}\left( \gamma _{l}-p_{m}\right) %
\right] ^{1/2}}{\prod_{\substack{ m=1  \\ m\neq l}}^{N}\left( \gamma
_{l}-\gamma _{m}\right) }\sum_{\substack{ m=1  \\ m\neq l}}^{N}\frac{1}{%
\gamma _{l}-\gamma _{m}}\left( \frac{\partial \gamma _{l}}{\partial
y^{\prime }}-\frac{\partial \gamma _{m}}{\partial y^{\prime }}\right)
\right\} \\
&&+4i\sum_{l=1}^{N}\left( -\frac{1}{2}\right) \frac{1}{\gamma _{l}^{3/2}}%
\left( \frac{\partial \gamma _{l}}{\partial y^{\prime }}\right) \frac{\left[
\prod_{m=1}^{2N}\left( \gamma _{l}-p_{m}\right) \right] ^{1/2}}{\prod 
_{\substack{ m=1  \\ m\neq l}}^{N}\left( \gamma _{l}-\gamma _{m}\right) }
\end{eqnarray*}
Using Eqs.(\ref{dgdxp}), (\ref{dgdyp}) and the conversion equation (\ref
{conv}) this expression is rewritten 
\begin{eqnarray*}
\frac{\partial ^{2}u}{\partial y^{\prime }\partial x^{\prime }}
&=&4i\sum_{l=1}^{N}\frac{1}{\sqrt{\gamma _{l}}}\left\{ \frac{1}{4i}\left( -%
\frac{\lambda ^{2}}{16}\right) \exp \left( u\right) \frac{1}{\gamma
_{l}^{3/2}}\left( \frac{\partial \gamma _{l}}{\partial x^{\prime }}\right)
^{2}\sum_{m=1}^{2N}\frac{1}{\left( \gamma _{l}-p_{m}\right) }\right. \\
&&\left. -\frac{1}{2i}\left( -\frac{\lambda ^{2}}{16}\right) \exp \left(
u\right) \frac{1}{\sqrt{\gamma _{l}}}\left( \frac{\partial \gamma _{l}}{%
\partial x^{\prime }}\right) \sum_{\substack{ m=1  \\ m\neq l}}^{N}\frac{1}{%
\gamma _{l}-\gamma _{m}}\left( \frac{1}{\gamma _{l}}\frac{\partial \gamma
_{l}}{\partial x^{\prime }}-\frac{1}{\gamma _{m}}\frac{\partial \gamma _{m}}{%
\partial x^{\prime }}\right) \right\} \\
&&+4i\sum_{l=1}^{N}\left( -\frac{1}{4i}\right) \left( -\frac{\lambda ^{2}}{16%
}\right) \exp \left( u\right) \frac{1}{\gamma _{l}^{3}}\left( \frac{\partial
\gamma _{l}}{\partial x^{\prime }}\right) ^{2}
\end{eqnarray*}
We finally obtain four terms in the expression 
\begin{equation}
\frac{\partial ^{2}u}{\partial y^{\prime }\partial x^{\prime }}=-\frac{%
\lambda ^{2}}{16}\exp \left( u\right) \left( T_{1}-2T_{2}+2T_{3}-T_{3}\right)
\label{d2uxpyp1}
\end{equation}
\begin{equation*}
T_{1}=\sum_{l=1}^{N}\frac{1}{\gamma _{l}^{2}}\left( \frac{\partial \gamma
_{l}}{\partial x^{\prime }}\right) ^{2}\sum_{m=1}^{2N}\frac{1}{\left( \gamma
_{l}-p_{m}\right) }
\end{equation*}
\begin{equation*}
T_{2}=\sum_{l=1}^{N}\frac{1}{\gamma _{l}^{2}}\left( \frac{\partial \gamma
_{l}}{\partial x^{\prime }}\right) ^{2}\sum_{\substack{ m=1  \\ m\neq l}}^{N}%
\frac{1}{\gamma _{l}-\gamma _{m}}
\end{equation*}
\begin{equation*}
T_{3}=\sum_{l=1}^{N}\sum_{\substack{ m=1  \\ m\neq l}}^{N}\frac{1}{\gamma
_{l}-\gamma _{m}}\frac{1}{\gamma _{l}\gamma _{m}}\frac{\partial \gamma _{l}}{%
\partial x^{\prime }}\frac{\partial \gamma _{m}}{\partial x^{\prime }}
\end{equation*}
\begin{equation*}
T_{4}=\sum_{l=1}^{N}\frac{1}{\gamma _{l}^{3}}\left( \frac{\partial \gamma
_{l}}{\partial x^{\prime }}\right) ^{2}
\end{equation*}

This expression must be compared with 
\begin{equation}
\frac{\partial ^{2}u}{\partial y^{\prime }\partial x^{\prime }}=\frac{%
\lambda ^{2}}{8}\left[ \exp \left( u\right) -1\right]  \label{d2uxpyp2}
\end{equation}

It can be verified that, for an arbitrary set of $p_{k},k=1,...2N$, and a
set of functions $\gamma _{l}\left( x,y\right) ,l=1,...,N$ verifying the
differential equations (\ref{dgammax}) and (\ref{dgammay}) (or,
equivalently, Eqs.(\ref{dgdxp}) and (\ref{dgdyp}) ) the two expressions (\ref
{d2uxpyp1}) and (\ref{d2uxpyp2}) are \emph{identical}. This means that the
initial nonlinear equation is verified if $u\left( x,y\right) $ is given by
the expression (\ref{expu}). The verification can be done by summing the
residuues in a formal expression defined by integration in the complex plane
of a function having an adequate singularity structure. We note however that
the expression can be verified also on purely algebraic grounds, chosing 
\emph{arbitrary} sets $\left\{ p_{k}\right\} $ and $\left\{ \gamma
_{k}\right\} $. From these numbers one calculates the derivatives appearing
in the four terms of the above formula, without any need to solve the
differential equations. The expression is verified simply as an algebraic
expression, by a symbolic software, or, for any particular choice the
verification can be done numerically.

We conclude that we dispose at this moment of a method to find a solution of
the Jacobs-Rebbi equation on a periodic spatial domain. This consists of
chosing a set of $2N$ arbitrary complex numbers, $\left\{ p_{k}\right\} $
and solving the first order differential equations for $\left\{ \gamma
_{k}\right\} $ with a set of initial conditions.

\subsection{Solving the equations for the auxiliary spectrum}

To solve the differential equations for $\gamma _{k}\left( x,y;p\right)
,k=1,...,N$ starting from a set of initial conditions is a difficult task as
is apparent from the form of the Eqs.(\ref{dgammax}) and (\ref{dgammay}).
However there is a standard procedure that provides the analytic solution of
these equations. It is based on the fundamental property of $\gamma
_{k}\left( x,y;p\right) $ of being defined when $p$ maps the complex plane
(of the spectral variable of the Lax operator) to the complex function given
by the square root of the Wronskian. Since the later is a polynomial in $p$,
the square root defines a \emph{hyperelliptic Riemann surface}, \emph{i.e.}
a compactified double covering of the complex plane with cuts connecting
pairs of zeros of the Wronskian. These are the points $\left\{
p_{k},k=1,...,2N\right\} $ of the main spectrum, plus the point zero and the
point at infinity. The point zero appears since in formulas (\ref{dgammax})
and (\ref{dgammay}) a factor of $\sqrt{\gamma _{k}}$ can be adjoined to the
product of the $l=1,...,2N$ differences $\left( \gamma _{k}-p_{l}\right) $,
simply by taking formally $p_{0}=0$. Then the object which can be defined on
the basis of the square root of the Wronskian but reflecting the need for
the particular form in the equations of $\gamma _{k}$ 's is 
\begin{equation*}
R\left( p\right) =\sqrt{\prod_{l=0}^{2N}\left( p-p_{k}\right) }
\end{equation*}
with $p_{0}=0$. The geometry of this hyperelliptic surface is important in
finding the solution.

Pairs of zeros $p_{k}$ are joined by cuts and in addition the origin is
connected to infinity. This gives a number of $N+1$ cuts and generates a
compact Riemann surface of genus $g=N$.

On this surface there are defined two objects characterising the
differential geometry of the curve:

\begin{itemize}
\item  a basis of the one dimensional cohomology group of the surface; this
means two sets each of $N$ closed paths on the curve (\emph{cycles}), having
particular intersection properties. The two sets are noted $a_{j}$, and
respectively $b_{j}$, $j=1,...,N$. The intersections are 
\begin{eqnarray*}
a_{j}\circ a_{k} &=&0 \\
a_{j}\circ b_{k} &=&\delta _{jk} \\
b_{j}\circ b_{k} &=&0
\end{eqnarray*}
A typical example, for an elliptic curve $g=1$ with the topology of the
torus, consists of the two possible closed turns around the torus, the short
way ($a$) and the long way ($b$).

\item  a basis in the ring of the one-dimensional differential forms 
\begin{equation*}
d\mu _{k}=\frac{p^{N-k}dp}{R\left( p\right) },\;k=1,...,N
\end{equation*}
\end{itemize}

With these two sets one calculate several quantities which are invariants of
the Riemann surface. Essentially there are calculated integrals of the
elements of the basis of differential forms along the cycles $a_{j}$ and $%
b_{j}$. These are called \emph{periods} and are organised in two matrices 
\begin{equation*}
A_{ij}=\int_{a_{j}}d\mu _{i}=\int_{a_{j}}\frac{p^{N-i}dp}{R\left( p\right) }%
\;,\;i=1,N,\;j=1,N
\end{equation*}
\begin{equation*}
B_{ij}=\int_{b_{j}}d\mu _{i}=\int_{b_{j}}\frac{p^{N-i}dp}{R\left( p\right) }%
\;,\;i=1,N,\;j=1,N
\end{equation*}
It is useful to work with the \emph{inverse} of the matrix $A$%
\begin{equation*}
C=A^{-1}
\end{equation*}
Using $C$, the matrix of $A$ periods is reduced at the identity matrix, and
the matrix $B$ becomes 
\begin{equation}
\tau =CB  \label{taud}
\end{equation}
the $\tau $-matrix, with positive imaginary part.

Using this geometrical framework the solution of the $\gamma _{k}$ equations
can be obtained by oparating first a transformation from the set $\left\{
\gamma _{k}\right\} $ to a set of functions $\left\{ \phi _{k}\right\} $
representing \emph{phases} of motion along the cycles of the Riemann
surface. This transformation effectively \emph{linearises} the motion, which
can be trivially integrated in these new variables.

We have to define the functions of the target set, the phases $\left\{ \phi
_{k}\right\} $. They are integrals of linear combinations of the
differential one-forms along paths on the Riemann surface, each starting
from an initial point $\gamma _{0}$ and ending in the point which correspond
to a function $\gamma _{l}$. The integrand is a combination of the
differential one-forms with coefficients from the matrix $C=A^{-1}$%
\begin{equation}
\phi _{k}=-\sum_{l=1}^{N}\int_{\gamma _{0}}^{\gamma
_{l}}\sum_{m=1}^{N}C_{km}d\mu _{m}  \label{Abelmap}
\end{equation}
The mapping that realises the correspondence from a collection of points $%
\left\{ \gamma _{l},l=1,N\right\} $ of the hyperelliptic Riemann surface to
a manifold defined by the collection of points $\left\{ \phi
_{k},k=1,N\right\} $ is called \emph{Abel map}. The manifold generated by
the points $\left\{ \phi _{k},k=1,N\right\} $ has genus $g=N$ (as the
initial curve) and has the topology of a torus. It is called \emph{Jacobi
torus}.

Since the upper limit in the integrals are precisely our points $\gamma _{k}$%
, we can obtain the differential equations for $l_{k}$ by direct derivation
of this formula and using the differential equations for $\gamma _{k}$. 
\begin{eqnarray*}
\frac{\partial \phi _{k}}{\partial x} &=&-\sum_{l=1}^{N}\frac{\partial
\gamma _{l}}{\partial x}\sum_{m=1}^{N}C_{km}d\mu _{m}\left( \gamma
_{l}\right) \\
&=&-\sum_{m=1}^{N}C_{km}\sum_{l=1}^{N}\frac{\gamma _{l}^{N-m}}{R\left(
\gamma _{l}\right) }\frac{\partial \gamma _{l}}{\partial x}
\end{eqnarray*}
and 
\begin{equation*}
\frac{\partial \phi _{k}}{\partial y}=-\sum_{m=1}^{N}C_{km}\sum_{l=1}^{N}%
\frac{\gamma _{l}^{N-m}}{R\left( \gamma _{l}\right) }\frac{\partial \gamma
_{l}}{\partial y}
\end{equation*}
Replacing the derivatives from Eqs.(\ref{dgammax}) and (\ref{dgammay}) we
have 
\begin{equation}
\frac{\partial \phi _{k}}{\partial x}=2\sum_{m=1}^{N}C_{km}\sum_{l=1}^{N}%
\frac{\gamma _{l}^{N-m}}{\prod_{\substack{ n=1  \\ n\neq l}}^{N}\left(
\gamma _{l}-\gamma _{n}\right) }\frac{1}{\sqrt{\gamma _{l}}}\left[ \frac{%
\lambda ^{2}}{16\sqrt{\gamma _{l}}}\frac{\left( \prod_{n=1}^{N}\gamma
_{n}\right) ^{2}}{\prod_{n=1}^{2N}p_{n}}+\sqrt{\gamma _{l}}\right]
\label{dphikx}
\end{equation}
and 
\begin{equation}
\frac{\partial \phi _{k}}{\partial y}=2i\sum_{m=1}^{N}C_{km}\sum_{l=1}^{N}%
\frac{\gamma _{l}^{N-m}}{\prod_{\substack{ n=1  \\ n\neq l}}^{N}\left(
\gamma _{l}-\gamma _{n}\right) }\frac{1}{\sqrt{\gamma _{l}}}\left[ \frac{%
\lambda ^{2}}{16\sqrt{\gamma _{l}}}\frac{\left( \prod_{n=1}^{N}\gamma
_{n}\right) ^{2}}{\prod_{n=1}^{2N}p_{n}}-\sqrt{\gamma _{l}}\right]
\label{dphiky}
\end{equation}

We have to calculate separately the two terms in each of the above formulas. 
\begin{equation*}
\rho _{1}\equiv \sum_{l=1}^{N}\frac{\gamma _{l}^{N-m-1}}{\prod_{\substack{ %
n=1  \\ n\neq l}}^{N}\left( \gamma _{l}-\gamma _{n}\right) }\frac{\lambda
^{2}}{16}\frac{\left( \prod_{n=1}^{N}\gamma _{n}\right) ^{2}}{%
\prod_{n=1}^{2N}p_{n}}
\end{equation*}
\begin{equation*}
\rho _{2}=\sum_{l=1}^{N}\frac{\gamma _{l}^{N-m}}{\prod_{\substack{ n=1  \\ %
n\neq l}}^{N}\left( \gamma _{l}-\gamma _{n}\right) }
\end{equation*}
Since the product of all the $\gamma _{n}$ 's is independent of the
summation index $l$, we will factorse it, as well as the product of the
eigenvalues $p_{n}$ and the constant 
\begin{equation*}
\rho _{1}=\frac{\lambda ^{2}}{16}\frac{\left( \prod_{n=1}^{N}\gamma
_{n}\right) ^{2}}{\prod_{n=1}^{2N}p_{n}}\sum_{l=1}^{N}\frac{\gamma
_{l}^{N-m-1}}{\prod_{\substack{ n=1  \\ n\neq l}}^{N}\left( \gamma
_{l}-\gamma _{n}\right) }
\end{equation*}
Tracy (for the case of Nonlinear Schrodinger Equation) \cite{Tracy} and Ting, Chen, Lee
(for \emph{sinh}-Poisson equation) \cite{TingChenLee} adopt different procedures to calculate
the sum. For example, one can write the Lagrange interpolation fromula for
an arbitrary function on a set of $N$ points $\left\{ x_{k}\right\} $%
\begin{equation*}
f\left( x\right) =\sum_{j=1}^{N}f\left( x_{j}\right) \frac{%
\prod_{k=1}^{N}\left( x-x_{k}\right) }{\prod_{\substack{ k=1  \\ k\neq j}}%
^{N}\left( x_{j}-x_{k}\right) }
\end{equation*}
Then one takes 
\begin{equation*}
f\left( x\right) \equiv x^{q}
\end{equation*}
then 
\begin{equation*}
x^{q}=\sum_{j=1}^{N}x_{j}^{q}\frac{\prod_{\substack{ k=1  \\ k\neq j}}%
^{N}\left( x-x_{k}\right) }{\prod_{\substack{ k=1  \\ k\neq j}}^{N}\left(
x_{j}-x_{k}\right) }
\end{equation*}
If the product at the numerator is expanded one gets a polynomial of degree $%
N$ while in the left hand side we have a polynomial of degree $q$. Comparing
the coefficients of the same powers of the variable $x$ in both sides it is
obtained 
\begin{equation*}
\delta \left[ q-(N-1)\right] =\sum_{j=1}^{N}x_{j}^{q}\frac{1}{\prod 
_{\substack{ k=1  \\ k\neq j}}^{N}\left( x_{j}-x_{k}\right) }
\end{equation*}
From this we find that 
\begin{eqnarray*}
\sum_{l=1}^{N}\frac{\gamma _{l}^{N-m-1}}{\prod_{\substack{ n=1  \\ n\neq l}}%
^{N}\left( \gamma _{l}-\gamma _{n}\right) } &=&\delta \left[ N-m-1-\left(
N-1\right) \right] \\
&=&\delta \left( m\right)
\end{eqnarray*}
and 
\begin{eqnarray*}
\sum_{l=1}^{N}\frac{\gamma _{l}^{N-m}}{\prod_{\substack{ n=1  \\ n\neq l}}%
^{N}\left( \gamma _{l}-\gamma _{n}\right) } &=&\delta \left[ N-m-\left(
N-1\right) \right] \\
&=&\delta \left( 1-m\right)
\end{eqnarray*}
\begin{equation*}
\rho _{1}=\frac{\lambda ^{2}}{16}\frac{\left( \prod_{n=1}^{N}\gamma
_{n}\right) ^{2}}{\prod_{n=1}^{2N}p_{n}}\delta \left( m\right)
\end{equation*}
\begin{equation*}
\rho _{2}=\delta \left( 1-m\right)
\end{equation*}
and the equations becomes 
\begin{eqnarray*}
\frac{\partial \phi _{k}}{\partial x} &=&2\sum_{m=1}^{N}C_{km}\left[ \frac{%
\lambda ^{2}}{16}\frac{\left( \prod_{n=1}^{N}\gamma _{n}\right) ^{2}}{%
\prod_{n=1}^{2N}p_{n}}\delta \left( m\right) +\delta \left( 1-m\right) %
\right] \\
&=&2C_{k1}
\end{eqnarray*}
\begin{eqnarray*}
\frac{\partial \phi _{k}}{\partial y} &=&2i\sum_{m=1}^{N}C_{km}\left[ \frac{%
\lambda ^{2}}{16}\frac{\left( \prod_{n=1}^{N}\gamma _{n}\right) ^{2}}{%
\prod_{n=1}^{2N}p_{n}}\delta \left( m\right) -\delta \left( 1-m\right) %
\right] \\
&=&-2iC_{k1}
\end{eqnarray*}

The equations can be trivially integrated and we obtain the $\left(
x,y\right) $ dependence of the phases 
\begin{equation}
\phi _{k}\left( x,y\right) =2C_{k1}\left( x-iy\right) +\phi _{k0}
\label{phik}
\end{equation}
where $\phi _{k0}$ are constants of integration, initial phases.

We note from Eq.(\ref{phik}) that the motion on the Jacobi torus is entirely
determined by the main spectrum through the topological properties of the
hyeprelliptic Riemann surface (canonical cycles, differential forms, period
matrices).

\subsection{The Jacobi inversion}

After the determination of the phases $\phi _{k}$, which are points on the
Jacobi torus, we want to be able to retrive the functions $\gamma _{k}\left(
x,y\right) $ of the auxiliary spectrum, since they are necessary for the
explicit determination of the solution $u\left( x,y\right) $, via Eq.(\ref
{expu}). This constitutes the Jacobi inversion problem and has been solved
in connection with elliptic functions. The main instrument is the Riemann $%
theta$ function.

The definition of the Riemann $theta$ function involves a vector of
dimension $N$ (we denote it by $\mathbf{\phi }$) and a $N\times N$ matrix $%
\tau $ whose elements have the imaginary part positive. 
\begin{equation*}
\Theta \left( \mathbf{\phi },\tau \right) =\sum_{m_{1}=-\infty }^{\infty
}\cdots \sum_{m_{n}=-\infty }^{\infty }\exp \left( 2\pi
i\sum_{k=1}^{N}m_{k}\phi _{k}+\pi i\sum_{i=1}^{N}\sum_{j=1}^{N}m_{i}\tau
_{ij}m_{j}\right)
\end{equation*}
In general the $\Theta $ function is associated to a hyperelliptic Riemann
surface of genus $N$ generated for example from a two sheeted covering of
the complex plane with $2N+1$ or $2N+2$ branch points, between which $N+1$
cuts ahve been done. The matrix $\tau $ corresponds to the matrix determined
from the periods of the canonical differential one-forms on the canonical
cycles, see Eq.(\ref{taud}). The argument of the $\Theta $ function is the
vector $\mathbf{\phi }$ . The following periodicity properties of the $%
\Theta $ functions are useful in the inversion problem:

\begin{enumerate}
\item  The translation with unity of only one component of the vector $%
\mathbf{\phi }$ leaves $\Theta $ invariant. This is actually related with
the fact that the components of the arguments $\phi _{k}$ are coordinates
along the cycles of the $g$-torus, and so they are periodical. Using the
symbol $\mathbf{e}_{k}$ for a column vector of $N$ components having only
one $1$ in position $k$ and $0$ in rest, we have 
\begin{equation*}
\Theta \left( \mathbf{\phi +e}_{k},\tau \right) =\Theta \left( \mathbf{\phi }%
,\tau \right)
\end{equation*}

\item  Adding to the argument $\mathbf{\phi }$ a vector consisting of one of
the columns (say, $j$) of the matrix $\tau $ generates a factor to the
function $\Theta $%
\begin{equation*}
\Theta \left( \mathbf{\phi +\tau }_{j},\tau \right) =\exp \left( 2\pi i\phi
_{j}-\pi i\tau _{jj}\right) \Theta \left( \mathbf{\phi },\tau \right)
\end{equation*}
\end{enumerate}

The function $\Theta $ with argument a vector of dimension $N$ has $N$
zero's. \emph{These roots of the }$\Theta $\emph{\ function solves the
Jacobi inversion problem}.

Certain necessary quantities must be defined. We consider again a linear
combination of the canonical differential one-forms $d\mu _{k}$ with
coefficients taken from the columns of the matrix $C=A^{-1}$. These linear
combinations are integrated on the Riemann surface along paths staring from
an arbitrary point $\gamma _{0}$ and ending in some point of the surface, $q$%
\begin{eqnarray}
\nu _{i}\left( q\right) &=&\int_{\gamma _{0}}^{q}d\nu _{i}=\int_{\gamma
_{0}}^{q}\sum_{m=1}^{N}C_{im}d\mu _{m}  \label{niuiq} \\
&=&\int_{\gamma _{0}}^{q}\sum_{m=1}^{N}C_{im}\frac{p^{N-m}dp}{R\left(
p\right) }  \notag
\end{eqnarray}
These are functions of the current point on the Riemann surface, $q$. We
consider the sum of the integrals of such functions along the $a$-cycles
plus terms from the diagonal of $\tau $%
\begin{equation}
D_{i}=-\frac{1}{2}\tau _{ii}+\sum_{j=1}^{N}\int_{a_{j}}\nu _{i}\left(
p\right) d\nu _{j}  \label{di}
\end{equation}
Finally, it is considered the function 
\begin{equation}
\zeta \left( q\right) =\Theta \left( \mathbf{\nu }\left( q\right) -\mathbf{%
\phi }+\mathbf{D},\tau \right)  \label{zetaq}
\end{equation}

It is proved that the zero's of the function $\zeta \left( q\right) $ are $%
\gamma _{k}$, the auxiliary spectrum.

\subsection{Solution of the Jacobs-Rebbi equation in terms of Riemann $%
\protect\theta $-functions}

Any initial condition for the nonlinear Jacobs-Rebbi equation leads to a
main spectrum, \emph{i.e.} a set of complex numbers $\left\{
p_{k},k=1,2N\right\} $. From these we construct the hyperelliptic Riemann
surface of genus $N$ and calculate the period matrices and the phases of the
linear motion along the canonical $a_{j}$ cycles on the surface. This is
purely topological and geometrical data, generated from the main spectrum
or, equivalently, by the initial condition for the unknown solution $u$.

On the other hand, solving the Jacobi inversion problem provides us with the
auxiliary spectrum $\left\{ \gamma _{k},k=1,N\right\} $ where $\gamma _{k}$
are functions of the phases, and, as such, of the variables $\left(
x,y\right) $.

The eigenvalues of the main spectrum and the functions $\gamma _{k}\left(
x,y\right) $ of the auxiliary spectrum give the explicit form of the
solution $u\left( x,y\right) $ via the conversion formula 
\begin{eqnarray}
u &=&\ln \left[ \frac{\left( \prod_{k=1}^{N}\gamma _{k}\right) ^{2}}{%
\prod_{m=1}^{2N}p_{m}}\right]  \label{usum} \\
&=&2\sum_{k=1}^{N}\ln \gamma _{k}-\sum_{m=1}^{2N}\ln p_{m}  \notag
\end{eqnarray}
Returning to the result of the Jacobi inversion procedure, we will try to
express the first sum in terms of the $\Theta $ function's zero's, \emph{i.e.%
} in terms of the zero's of the function $\zeta \left( q\right) $.

In Ting, Chen and Lee \cite{TingChenLee} it is adopted the method consisting of generating
directly the sum of the logarithms from an integral of a complex function.

We have to remind that the hyperelliptic Riemann surface is a mapping from
the complex plane of the spectral parameter $p$ via the square root of the
polynomial expression generated by the Wronskian. The variable $q$ appearing
as the upper limit of integration in Eq.(\ref{niuiq}) is a point on the
hyperelliptic Riemann surface and is the image of a point in the complex $p$%
-plane; the path of integration in Eq.(\ref{niuiq}) is the image of a path
on the same $p$-plane. We can try to introduce an intermediate object, a
complex function whose singularities will lead us, after integration, to the
sum of logarithms. This is 
\begin{equation}
w=\frac{1}{2\pi i}\int_{\Gamma }f\left( q\right) \frac{d\zeta \left(
q\right) }{\zeta \left( q\right) }  \label{dzeta}
\end{equation}
with the contour of integration $\Gamma $ being a path on the Riemann
surface. This path must be chosen such that it circles all the zero's of $%
\zeta \left( q\right) $, and then the value of the integral will be 
\begin{equation*}
w=\sum_{j=1}^{N}f\left( q_{0j}\right)
\end{equation*}
where $q_{0j},j=1,N$ are 
\begin{equation*}
\zeta \left( q_{0j}\right) =0
\end{equation*}
The contour $\Gamma $ is specified after the Riemann surface is mapped back
onto the $p$-plane as the normal polygon obtained from cutting along the
canonical $a_{j}$ and $b_{j}$ cycles. Since the genus of the Riemann surface
is $N$, the number of cycles is $2N$ and each cycle generates two edges of
the polygon, with opposite senses. The polygon has $4N$ edges and it is
chosen as the contour $\Gamma $. All the points $q_{0j}$ are somewhere
inside the polygon, so what we need is a choice for the function $f\left(
q\right) $. It is natural to take 
\begin{equation*}
f\left( q\right) \equiv \ln q
\end{equation*}
since we want the sum of the logarithms, but this induces an additional
singularity at $q=0$ and a cut connecting $q=0$ to $\infty $ on the $p$%
-plane. This cut on the $p$-plane is translated into \emph{two} paths on the
hyperelliptic Riemann surface and since the variable of integration on the
path is $\zeta $ we have to connect the two points into which $p=0$ is
mapped with the single point on the surface that corresponds to $p=\infty $.
This actually separates the polygon $\Gamma $ into two closed parts. The
integration in Eq.(\ref{dzeta}) must be done separately on the two contours.
A part of the integration will be done along the cut and in one integration
the path correspond to one determination of the logarithm (one branch) while
in the other integration the path is on the next branch of the logarithm. 
\begin{eqnarray*}
J_{+} &=&-\int_{0}^{\infty }\ln q\frac{d\zeta }{\zeta }+\int_{0}^{\infty
}\left( \ln q+2\pi i\right) \frac{d\zeta }{\zeta } \\
&=&2\pi i\int_{0}^{\infty }\frac{d\zeta }{\zeta }
\end{eqnarray*}
\begin{eqnarray*}
J_{-} &=&-\int_{0}^{\infty }\left( \ln q+2\pi i\right) \frac{d\zeta }{\zeta }%
+\int_{0}^{\infty }\left( \ln q+4\pi i\right) \frac{d\zeta }{\zeta } \\
&=&2\pi i\int_{0}^{\infty }\frac{d\zeta }{\zeta }
\end{eqnarray*}
The rest of the integration is the sum over the poles of the integrand, 
\emph{i.e.} the zero's of the function $\zeta \left( q\right) $%
\begin{equation}
\frac{1}{2\pi i}\int_{\Gamma }\ln q\frac{d\zeta }{\zeta }=\sum_{j=1}^{N}\ln
\gamma _{j}-2\int_{0}^{\infty }\frac{d\zeta }{\zeta }  \label{intga}
\end{equation}

These are two images of the compactified two sheeted covering of the complex
plane in two Riemann hyperelliptic curves with genus $2$ and respectively
genus $3$. They correspond with the case where the number of branch points
in the main spectrum $p_{k}$ is $2g+2=6$ and respectively $8$ , \emph{i.e.}
if the eigenvalues comes in three or four pairs.
\begin{figure}[htbp]
\centerline{\includegraphics[height=7cm]{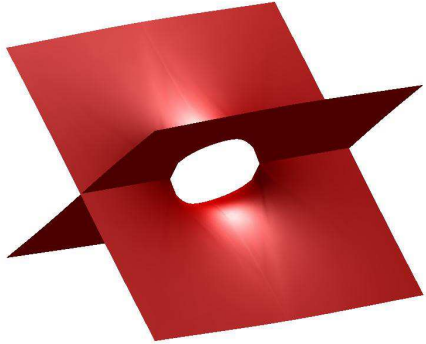}}
\caption{Two sheet Riemann surface of $y=\sqrt{x^2-1}$.}
\label{fig1}
\end{figure}
\begin{figure}[htbp]
\centerline{\includegraphics[height=7cm]{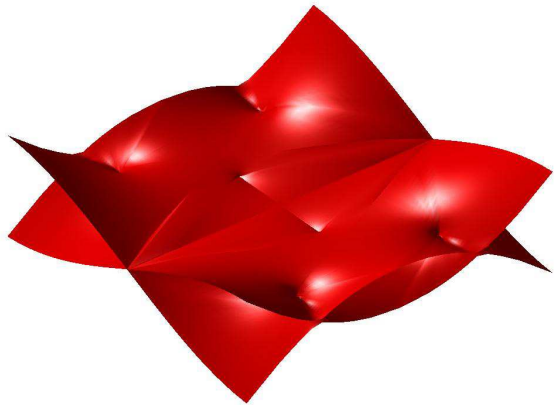}}
\caption{Branched covering of the the complex plane by the
two sheet Riemann surface of the function 
$y=\sqrt{\left(z^2+a\right)\left(z^2+b\right)\left(z^2+c\right)}$.
This would correspond to a main spectrum consisting of only $n=6$ points.}
\label{fig2}
\end{figure}

\begin{figure}[htbp]
\centerline{\includegraphics[height=7cm]{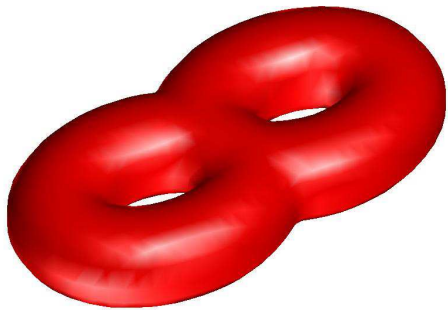}}
\caption{The hyperelliptic Riemann surface of genus $g=2$. 
This is topologically equivalent to the two-sheet Riemann
surface shown in \ref{fig2} since $n=2g+2=6$ gives $g=2$.}
\label{fig3}
\end{figure}

\begin{figure}[htbp]
\centerline{\includegraphics[height=7cm]{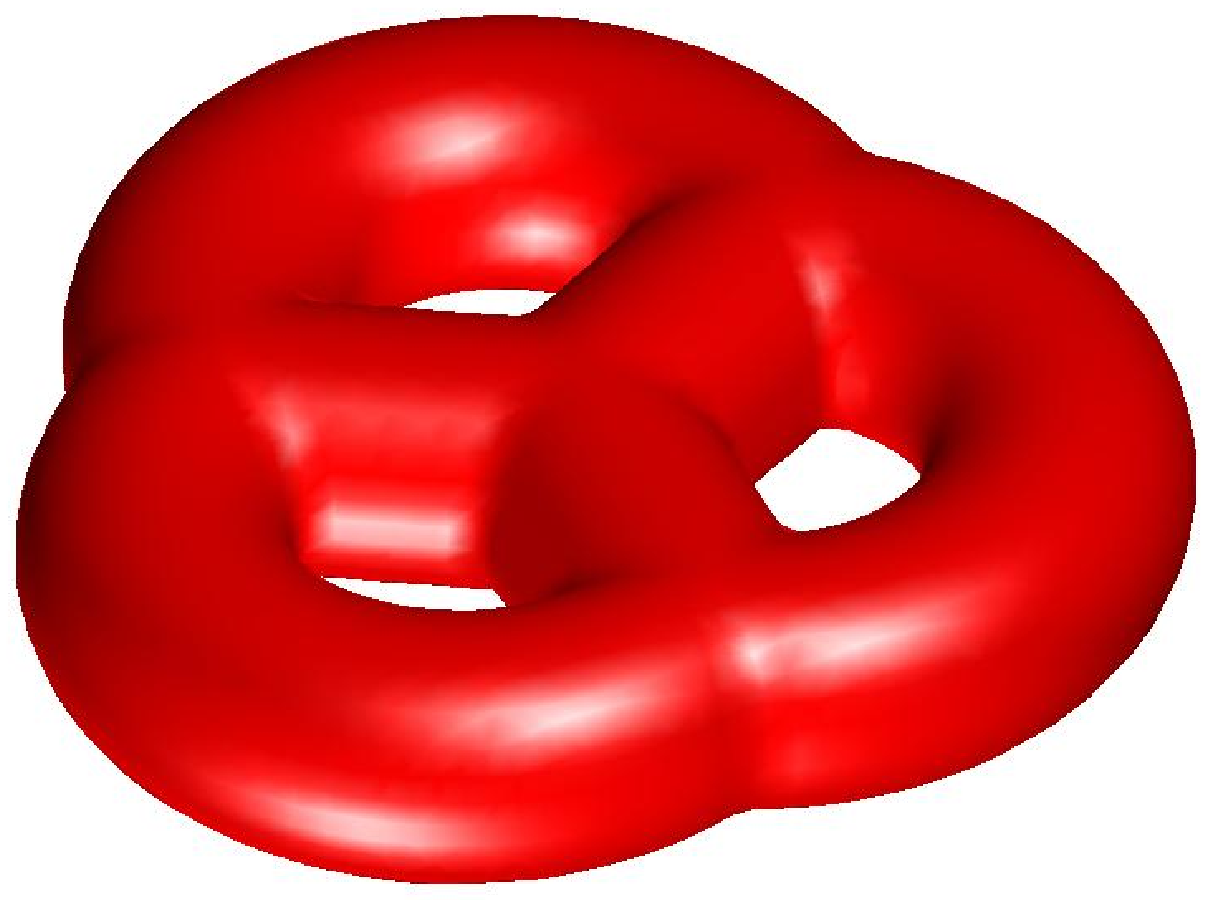}}
\caption{A hyperelliptic Riemann surface of $g=3$.}
\label{fig4}
\end{figure}

An alternative calculation of the same integral Eq.(\ref{dzeta}) is done
following directly the path along the canonical cycles $a$ and $b$. In this
evaluation the periodicity properties of the $\Theta $ function are
essential. When the point of integration $q$ is on a $b_{k}$ cycle, the
point that corresponds to it but attached to the opposite side of the cut
along the cycle can be reached by a complete turn along the nearest $a$
cycle. But such a change does not introduce any modification in the
integrand, since it is exactly the operation involved in the first
periodicity property of $\Theta $. This means that the integration along the
edge of $\Gamma $ coming from a cycle $b_{k}$ can be paired with the
integration along the edge coming from the opposite side of the cut along $%
b_{k}$, without any change in the integrand. Since these two integrals are
equal but of opposite sign we conclude that the edges originated from $b$%
-cycles do not contribute to the integral. Analoguous consideration for the $%
a_{k}$-cycles involve the second property of the $\Theta $ function. If $q$
is on an edge representing one side of the cut along the $a_{k}$ cycle, the
point $\overline{q}$ on other side can be reached by a complete turn along a 
$b$ cycle. This introduces the change of the integrand 
\begin{equation*}
\zeta \left( q\right) \rightarrow \zeta \left( \overline{q}\right) =\exp
\left\{ -i\pi \tau _{kk}-2\pi i\left[ \nu _{k}\left( q\right) -\phi
_{k}+D_{k}\right] \right\} \zeta \left( q\right)
\end{equation*}
This means 
\begin{eqnarray*}
\ln \zeta \left( \overline{q}\right) &=&-i\pi \tau _{kk}-2\pi i\left[ \nu
_{k}\left( q\right) -\phi _{k}+D_{k}\right] \\
&&+\ln \zeta \left( q\right)
\end{eqnarray*}
and 
\begin{eqnarray*}
d\ln \zeta \left( \overline{q}\right) &=&\frac{d\zeta \left( \overline{q}%
\right) }{\zeta } \\
&=&-2\pi id\nu _{k}\left( q\right) +\ln \zeta \left( q\right)
\end{eqnarray*}
Taking into account the conclusion reached before that the $b$ cycles do not
contribute to the integration and leaving aside the part coming from the
branch cut the integral becomes 
\begin{eqnarray*}
&&\frac{1}{2\pi i}\int_{\Gamma }\ln q\frac{d\zeta }{\zeta } \\
&=&\sum_{k=1}^{N}\left[ \int_{a_{k}}\ln q\frac{d\zeta \left( q\right) }{%
\zeta }+\int_{-a_{k}}\ln \overline{q}\frac{d\zeta \left( \overline{q}\right) 
}{\zeta }\right] \\
&=&\sum_{k=1}^{N}\int_{a_{k}}\left\{ \ln q\frac{d\zeta \left( q\right) }{%
\zeta }-\ln q\left[ 2\pi id\nu _{k}\left( q\right) +\frac{d\zeta \left(
q\right) }{\zeta }\right] \right\}
\end{eqnarray*}
or 
\begin{equation}
\frac{1}{2\pi i}\int_{\Gamma }\ln q\frac{d\zeta }{\zeta }=\sum_{k=1}^{N}%
\int_{a_{k}}\ln qd\nu _{k}  \label{intcons}
\end{equation}
This integral is a constant since $d\nu _{k}$ is a differential one-form
generated from a linear combination of the canonical one-forms (depending
only on the surface) and the integration is performed over closed loops $%
a_{k}$. It does not leave any choice since it does not depend on any
parameter.

We have completed the calculation of the integral (\ref{intga}) in the two
ways: one with the polygonal dissection of the Riemann surface plus the
branch cut (which gives the right hand side of (\ref{intga}) ) and one with
the path on the surface, using the reunion of canonical cycles, obtaining
the constant of Eq.(\ref{intcons}). It only remains to make explicit the
last term in Eq.(\ref{intga}) coming from the branch cut integration. 
\begin{equation}
\int_{0}^{\infty }\frac{d\zeta \left( q\right) }{\zeta }=\ln \frac{\zeta
\left( \infty \right) }{\zeta \left( 0\right) }  \label{ettw}
\end{equation}
with the relation 
\begin{eqnarray*}
\zeta \left( \infty \right) &=&\Theta \left[ \mathbf{\phi +\nu }\left(
\infty \right) -\mathbf{D}\right] \\
&=&\Theta \left[ \mathbf{\phi +\nu }\left( 0\right) +\int_{0}^{\infty }d%
\mathbf{\nu }\left( q\right) -\mathbf{D}\right]
\end{eqnarray*}
In the argument, $\mathbf{\nu }\left( 0\right) -\mathbf{D}$ is a constant
that can be included in the initial phases $\phi _{k0}$ (Eq.(\ref{phik})).
The integrals of the differential forms are done along a path that can be
completed with a circle at infinity. It results a loop can then be mapped
onto the set of loops that surround the cuts, \emph{i.e.} effectively it is
shrinked the set of $a$-cycles. The integrals are then reduced at the
diagonal entries of the $A$ matrix which are all unity. 
\begin{equation*}
\int_{0}^{\infty }d\mathbf{\nu }\left( q\right) =\frac{1}{2}\int_{C}d\mathbf{%
\nu }\left( q\right) =\frac{1}{2}\mathbf{I}
\end{equation*}
Then Eq.(\ref{ettw}) can be written 
\begin{equation*}
\int_{0}^{\infty }\frac{d\zeta \left( q\right) }{\zeta }=\ln \left[ \frac{%
\Theta \left( \mathbf{\phi +}\frac{1}{2}\mathbf{I}\right) }{\Theta \left( 
\mathbf{\phi }\right) }\right]
\end{equation*}

We return now to the Eq.(\ref{intga}) and (\ref{intcons}) 
\begin{eqnarray*}
\frac{1}{2\pi i}\int_{\Gamma }\ln q\frac{d\zeta }{\zeta } &=&\sum_{j=1}^{N}%
\ln \gamma _{j}-2\int_{0}^{\infty }\frac{d\zeta }{\zeta } \\
&=&\sum_{k=1}^{N}\int_{a_{k}}\ln qd\nu _{k}
\end{eqnarray*}
where from we obtain 
\begin{equation*}
\sum_{j=1}^{N}\ln \gamma _{j}=2\int_{0}^{\infty }\frac{d\zeta }{\zeta }%
+\sum_{k=1}^{N}\int_{a_{k}}\ln qd\nu _{k}
\end{equation*}

The explicit form of the solution is given by the conversion formula (\ref
{usum}) 
\begin{eqnarray*}
u &=&2\sum_{k=1}^{N}\ln \gamma _{k}-\sum_{m=1}^{2N}\ln p_{m} \\
&=&4\ln \left[ \frac{\Theta \left( \mathbf{\phi +}\frac{1}{2}\mathbf{I}%
\right) }{\Theta \left( \mathbf{\phi }\right) }\right] \\
&&+\sum_{k=1}^{N}\int_{a_{k}}\ln qd\nu _{k}-\sum_{m=1}^{2N}\ln p_{m}
\end{eqnarray*}
Since the last line is composed of constants, 
\begin{equation*}
K\equiv \sum_{k=1}^{N}\int_{a_{k}}\ln qd\nu _{k}-\sum_{m=1}^{2N}\ln p_{m}
\end{equation*}
we can write the solution as 
\begin{equation*}
u\left( x,y\right) =4\ln \left[ \frac{\Theta \left( \mathbf{\phi +}\frac{1}{2%
}\mathbf{I}\right) }{\Theta \left( \mathbf{\phi }\right) }\right] +K
\end{equation*}
with 
\begin{equation*}
\phi _{k}\left( x,y\right) =2C_{k1}\left( x-iy\right) +\phi _{k0}
\end{equation*}

\begin{figure}[htbp]
\centerline{\includegraphics[height=7cm]{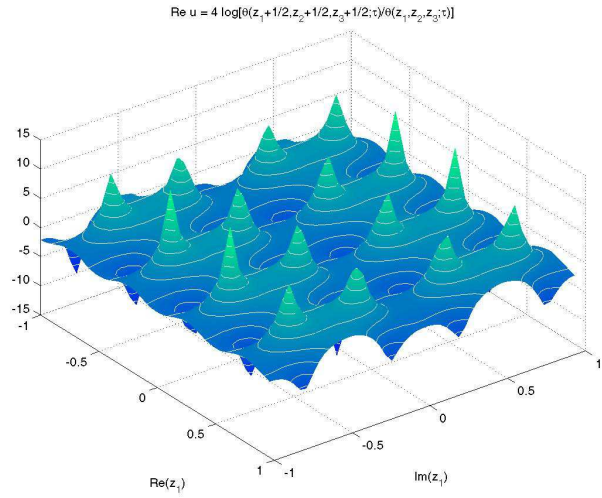}}
\caption{The solution of the Jacobs-Rebbi equation.}
\label{fig5}
\end{figure}

\begin{figure}[htbp]
\centerline{\includegraphics[height=7cm]{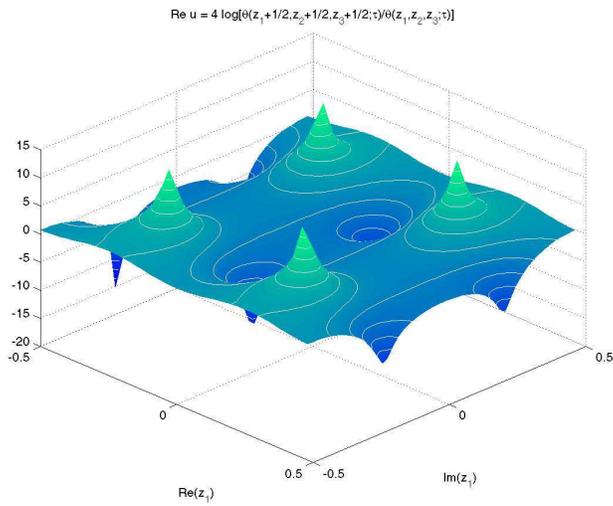}}
\caption{The solution clearly shows the vortical structures
as expected.}
\label{fig6}
\end{figure}

\section{Conclusion}

In conclusion we have proved that the Jacobs-Rebbi equation is exactly integrable and have provided
the exact solution. We have followed the standard approaches developed in detail for similar equations:
 \emph{sine}-Gordon and \emph{sinh}-Poisson equations. 

Knowledge of the exact solution will make more accessible the investigation of the physical applications
of this equation.

\end{document}